\documentclass{iaus}
\usepackage{graphicx}
\usepackage{url}

\title[NBursts+phot spectrophotometric inversion] 
{NBursts+phot:
parametric recovery of galaxy star formation histories from the simultaneous
fitting of spectra and broad-band spectral energy distributions}

\author[Chilingarian \& Katkov]   
{Igor V. Chilingarian$^1$\thanks{Present address: CfA/SAO 60 Garden St.,
Cambridge, MA, 02138 USA}
 \and Ivan Yu. Katkov$^1$}

\affiliation{$^1$Sternberg Astronomical Institute, Moscow State University, \\ 
Universitetskii pr. 13, Moscow, 119992 Russia 
\\ email {\tt IC: chil@sai.msu.ru; IK: katkov.ivan@gmail.com} \\
[\affilskip]}

\pubyear{2011} 
\volume{284}  
\pagerange{1--12}
\jname{The Spectral Energy Distribution of Galaxies}
\editors{R.J. Tuffs \&  C.C.Popescu, eds.}
\begin{document}

\maketitle

\begin{abstract} 
We present \textsc{NBursts+phot}, a novel technique for the parametric inversion of
spectrophotometric data for unresolved stellar populations where high-resolution 
spectra and broadband SEDs are fitted simultaneously helping to break the 
degeneracies between parameters of multi-component stellar population models.
\end{abstract}

\firstsection 
\section{Motivation}

Optical spectra of galaxies contain important information about their
internal kinematics and stellar content that can be extracted by fitting
them against stellar population models. There is a number of approaches for                                     
the full spectrum fitting that were proved to be efficient in studying
different classes of galaxies and star clusters.

On the other hand, the broad-band photometry in different spectral domains
that became available for large samples of galaxies thanks to modern
wide-field surveys allows one to study certain properties that cannot be
derived from the spectra (e.g. internal extinction) and to brake
degeneracies between parameters in case of complex star formation histories
(SFH). Usually, photometric measurements and spectra are used independently.
Here we propose a new approach \textsc{NBursts+phot} that fits in a single minimization
loop spectral and photometric information and recovers both, parametric SFH,
and internal kinematics of galaxies.

\section{Novel Technique} 

\textsc{NBursts+phot} is the extension of the \textsc{NBursts} (Chilingarian et al.
2007a,b) parametric full spectrum fitting
technique implemented quite similarly to the non-parametric spectrophotometric
inversion proposed by Pappalardo et al. (2010). During the minimization, the
$\chi^2$ value is computed as a sum of the spectral and photometric contributions
taken with a certain weight $\alpha$:
\begin{eqnarray}
        \chi^2 = \sum_{N_{\lambda}}\frac{(F_{i}-P_{1p}(T_{i}(SFH) \otimes
        \mathcal{L}(v,\sigma,h_3,h_4) + P_{2q}) )^2}{\Delta F_{i}^2} +
        \alpha \sum_{N_{ph}}\frac{(f_{j}-w_{j}t_{j}(SFH)A_{j})^2}{\Delta f_{j}^2}, 
        \nonumber\\
        \mbox{where} \quad T_{i}(SFH) = \sum_{N_{bursts}}k_{i} T_{i}(t_n,
        Z_n) \hskip 1cm
\label{chi2eq}
\end{eqnarray}

\noindent where $\mathcal{L}$ is LOSVD; $F_{i}$ and $\Delta F_{i}$ are
observed spectral flux and its uncertainty; $T_{i}(SFH)$ is the flux from an
synthetic spectrum, represented by a linear combination of $N_{bursts}$
SSP's and convolved according to the line-spread function of the
spectrograph; $P_{1p}$ and $P_{2q}$ are multiplicative and additive Legendre
polynomials of orders $p$ and $q$ for correcting the continuum; $t$ is age,
$Z$ is metallicity, $v$, $\sigma$, $h_3$, and $h_4$ are radial velocity,
velocity dispersion and Gauss-Hermite coefficients respectively (van der
Marel \& Franx, 1993); $f_{j}$ and $t_{j}$ are observed and modelled
photometric fluxes; $w_{j}$ is a photometric flux normalization coefficient
computed linearly; $A_{j}(E(B-V))$ is a photometric extinction in a given
band depending on $E(B-V)$, an additional free parameter. By adjusting the
$\alpha$ parameter, one can choose the relative importance of photometric
and spectral measurements. When dealing with galaxies at non-zero redshifts,
the photometric measurements should be $k$-corrected.

\begin{figure}
\begin{center}
 \includegraphics[width=\textwidth]{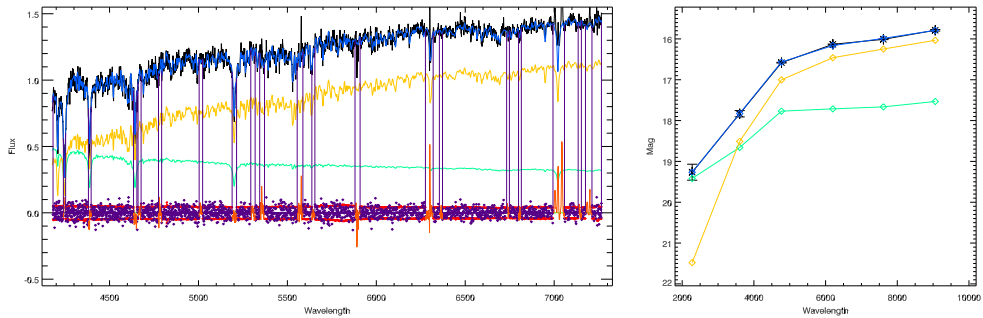} 
 \caption{Example of the spectrophotometric inversion for a poststarburst
 E+A galaxy SDSS~J230743.41+152558.4. The SDSS spectrum is shown in the left
 panel; the SED including GALEX NUV and 5 SDSS photometric measurements is
 shown in the right panel. Data (both, spectral and photometric) are shown
 in black, best-fitting model in blue and its two components in orange and
 green.}
   \label{fig_EpA}
\end{center}
\end{figure}

In Fig.~\ref{fig_EpA} we present an application of the \textsc{NBursts+phot}
technique to the combined dataset for a poststarburst E+A galaxy
SDSS~J230743.41+152558.4 (\cite{CDRB09}). We used its SDSS DR7
(\cite{SDSS_DR7}) spectrum
(left panel) and photometry from SDSS DR7 (5 optical $ugriz$ bands) and
GALEX GR5 (one $NUV$ band, \cite{Martin+05}) surveys. Photometric measurements were
$k$-corrected using the analytical approximations from \cite{CMZ10} and
\cite{CZ11} for optical and GALEX bands respectively\footnote{See
the $k$-corrections calculator at \url{http://kcor.sai.msu.ru/UVtoNIR.html}}.

\section{Pros and Cons}

\subsection{Pros}
\begin{itemize}
\item Photometry and spectra are fitted in a single loop so all the available
information is used to constraint the SFH and internal kinematics.
Therefore, certain biases originating from degeneracies between parameters
of internal kinematics and stellar populations (such as metallicity -
velocity dispersion) can be avoided.

\item Photometry in certain bands helps to break degeneracies between parameters
of multiple star formation episodes. For example, adding GALEX UV colours 
will help to detect small quantities of young stars and measure their mass
fraction, while spectra will be helpful to recover their parameters (age and
metallicity) which cannot be derived from photometric data alone.

\item We can estimate the internal extinction in galaxies while constraining their
stellar population properties by the spectral data.
\end{itemize}

\subsection{Cons}
\begin{itemize}
\item The choice of $\alpha$ is not evident. It has to be determined for every data
collection.

\item Spectral and photometric models have different origin and are
constructed from different ingredients, therefore the procedure is not fully
self-consistent and some biases can be introduced. For the spectral fitting
we use high-resolution SSP models computed with the PEGASE.HR
(\cite{LeBorgne+04}) code based on the ELODIE.3.1 (\cite{PSKLB07}) empirical
stellar library, while the photometric models are computed with the PEGASE.2
(\cite{FR97}) code using the low-resolution BaSeL synthetic stellar library
(\cite{LCB97}). Spectra of real galaxies often contain emission lines which
are sometimes quite strong and may influence the corresponding photometric
measurements.

\item It is difficult to get homogeneous photometry in several spectral domains
because the data provided by different wide-field surveys often have certain
specific features different, such as apertures, photometric zero points, in 
addition, the photometry has to be provided for the same aperture as a
spectrum (e.g. we cannot use the integrated galaxy photometry and 3-arcsec
SDSS spectra).
\end{itemize}

\bigskip

Authors thank the IAU for the provided financial aid and RFBR grant
10-02-00062a for covering the remaining expanses.

\end{document}